\documentclass[10pt,twocolumn,letterpaper]{article}

\usepackage{cvpr}
\usepackage{times}
\usepackage{epsfig}
\usepackage{graphicx}
\usepackage{amsmath}
\usepackage{amssymb}
\usepackage{verbatimbox}
\usepackage{algorithm} 
\usepackage{algpseudocode} 
\usepackage[normalem]{ulem}
\usepackage{amsmath}
\usepackage{cite}
\newcommand{\stkout}[1]{\ifmmode\text{\sout{\ensuremath{#1}}}\else\sout{#1}\fi}

\usepackage[breaklinks=true,
            bookmarks=false,
            colorlinks = true,
            linkcolor = blue,
            urlcolor  = blue,
            citecolor = black,
            anchorcolor = blue]{hyperref}

\cvprfinalcopy 

\def\cvprPaperID{****} 

\begin{document}

\title{Efficient Neural Network Deployment for Microcontroller}

\author{Hasan Unlu\\
Stanford University\\
450 Serra Mall, Stanford, CA 94305\\
{\tt\small hunlu@stanford.edu}
}

\maketitle

\begin{abstract}
Edge computing for neural networks is getting important especially for low power applications and offline devices. TensorFlow Lite and PyTorch Mobile were released for this purpose. But they mainly support mobile devices instead of microcontroller level yet. Microcontroller support is an emerging area now. There are many approaches to reduce network size and compute load like pruning, binarization  \cite{Alizadeh} and layer manipulation i.e. operator reordering  \cite{Liberis}. This paper is going to explore and generalize convolution neural network deployment for microcontrollers with two novel optimization proposals offering memory saving and compute efficiency in 2D convolutions as well as fully connected layers. The first one is in-place max-pooling, if the stride is greater than or equal to pooling kernel size. The second optimization is to use ping-pong buffers between layers to reduce memory consumption significantly. The memory savings and performance will be compared with CMSIS-NN framework ~\cite{lai2018cmsisnn} developed for ARM Cortex-M CPUs. The final purpose is to develop a tool consuming PyTorch model with trained network weights, and it turns into an optimized inference engine(forward pass) in C/C++ for low memory(kilobyte level) and limited computing capable microcontrollers.
\end{abstract}

\section{Introduction}
Neural networks are mainly operated in GPUs or multi-core CPUs to increase throughput. Because any neural network architecture can be converted into data pipeline between layers. So each layer can be executed in different hardware sources in parallel. In addition to that, convolution kernels and dot products are perfectly suitable for parallel execution. Nowadays, neural networks are wanted to be run in low power systems and edge computing devices. These group of devices are primarily microcontrollers. However, they are usually single core with low data memory and executions happen sequentially. The sequential execution needs less memory than parallel execution, because only one layer or block of the neural network can run in per operation. So memories will be reusable between layers. The motivation of the ping-pong buffering in between layers is coming from this reusability. The max-pooling layer reduces output height and width of the input. Given certain condition of stride and kernel size of the pooling layer can reduce output buffer usage and it max-pools the set of the element without holding them in the interim buffer. This optimization improved to fuse max-pool layer into convolution output altogether with activation functions.
There is an effort implementing convolution, max-pooling, matrix product and activation functions to ARM Cortex-M microcontrollers getting help of parallelism of Multiply-and-Accumulate instruction in ARM Cortex-M4 and Cortex-M7 family ~\cite{lai2018cmsisnn}.

\section{Related Work}
Since the microcontroller deployment for neural network is recently emerging area, there are not many paper or work to compare. The closest paper ~\cite{lai2018cmsisnn} which also offers neural network framework for ARM Cortex-M4 and Cortex-M7 architecture is going to be reviewed. The paper implements convolution, max-pooling, dot product and activation functions for ARM Cortex-M4 and Cortex-M7 architectures. They are mainly focusing on performance and memory optimization. Since Cortex-M4 and Cortex-M7 have Multiply-and-Accumulate(MAC) instruction, it can execute two multiplications and summation of them in single instruction, this gives major performance advantage on matrix multiplication. They are reusing big scratch buffer between layers which is roughly similar to the ping-pong approach proposing in this paper except explicitly formulating upper memory limit and tight usage. They mention in-place max-pooling for just layer itself(max-pooling layer input and output) not fusing in convolution output to use final buffer size instead of output of convolution layer buffers. In their example, the implemented architecture is three convolution layers and one fully connected layer on CIFAR-10 dataset ~\cite{CIFAR}. The architecture is also quantized to \textit{int8} instead of 32-bit floating point(FP-32). Same architecture will be implemented and memory footprint will be compared.

\section{Optimization Methods}
To be able to show optimization results in numbers, LeNeT-5 architecture is selected and FP-32 is used in the implementation. LeNet-5 architecture(shown in Figure 1) was trained on MNIST ~\cite{MNIST} dataset. Loss functions is cross entropy loss and optimizer is Adam with learning rate of 2e-3. The model giving best accuracy on test set in 4 epochs was selected as best network to deploy. The following list shows the architecture parameters.

\begin{verbnobox}[\fontsize{7pt}{7pt}\selectfont]
The test set accuracy is 0.9844
(0): Conv2d(1, 6, kernel_size=(5, 5), stride=(1, 1))
(1): ReLU()
(2): MaxPool2d(kernel_size=2, stride=2, padding=0)
(3): Conv2d(6, 16, kernel_size=(5, 5), stride=(1, 1))
(4): ReLU()
(5): MaxPool2d(kernel_size=2, stride=2, padding=0)
(6): Flatten()
(7): Linear(in_features=400, out_features=120, bias=True)
(8): ReLU()
(9): Linear(in_features=120, out_features=84, bias=True)
(10): ReLU()
(11): Linear(in_features=84, out_features=10, bias=True)
\end{verbnobox}
The total parameter count here is $1*6*5*5+6+6*16*5*5+16+16*5*5*400+400+400*120+120+120*84+84+84*10+10=61706*sizeof(float)=246824$ bytes. ReLU layer can be part of the convolution layer, so there is no additional memory needed for it. The total memory usage based on caching between layers would be $32*32(input)+6*28*28(conv1)+6*14*14(maxpool1)+16*10*10(conv2)+16*5*5(maxpool2)+120(fc1)+84(fc2)+10(output)=9118*sizeof(float)=36472$ bytes. For easy calculation, float size can be assumed 4 bytes. Total memory requirement will be the total size of parameters and buffers between layers. It is $246824+36472=283296$ bytes which is $\sim$283 KBytes of memory. We can reduce this usage significantly using ping-pong buffers between layers, in-place max-pool and defining parameters read-only that moves all parameters to \textit{.text} region(non-volatile memory).
\begin{figure}[t]
\begin{center}
   \includegraphics[scale=0.17]{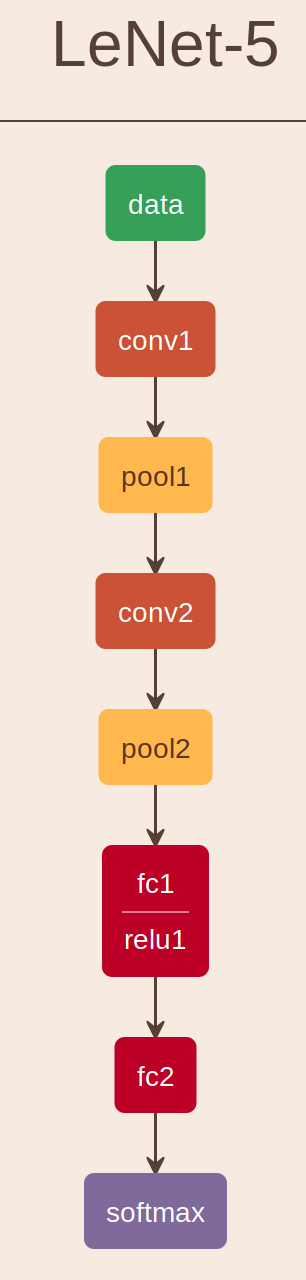}
\end{center}
   \caption{LeNet-5 Architecture ~\cite{LeCun}}
\label{fig:long}
\label{fig:onecol}
\end{figure}

\subsection{Fused in-place max-pooling in convolution layer}
This optimization will allow us to prevent additional for loops to execute over the convolution layer output for max-pooling and to use reduced output size instead of full size of the convolution output. If stride size of the max pooling layer is greater than or equal to size of the max-pooling kernel, we can calculate output while calculating convolution for max pooling kernel size without any cost of memory. Because, we do not need to keep all convolution results of max-pooling window in the memory, they are not going to be used again when stride is grater than or equal to max-pooling kernel. The stride will shift calculation to mutually exclusive new max-pooling block. Thus, we can reuse the existing intermediate buffer again for max-pooling output instead of creating new buffer. When stride condition meets, max-pooling output can be directly written into input buffer. Convolution layer output is not need to be stored before max-pool size. Peak memory usage between these two layer(convolution and max-pooling) will be $m*n/s^2$ instead of the expected output size $m*n$. Figure 2 shows two max-pooling iterations.

\begin{algorithm}
    \caption{Efficient 2D convolution with in-place max pooling when \textbf{max pooling kernel size is greater than or equal to stride}.\newline
    $m, n$ is dimension of the input.\newline
    $K$ is kernel coefficients.\newline
    $k$ is convolution kernel size\newline
    $s$ is stride of maxpooling layer and at the same time equals to kernel size of maxpooling layer} 
	\begin{algorithmic}[0]
	\For {$(x,y)=(0,0),(0,s),(0,2s)...(m, n) \ldots$}
      \State $max\_pooling\_element \leftarrow 0$
      \For {$(i,j)=(0,1),(0,2),(0,3)...(s-1, s-1) \ldots$}
	    \State $sum \leftarrow 0$
			    \For {$(z,t)=(0,0),(0,1),(0,2)...(k, k) \ldots$}
				    \State $sum \leftarrow sum + in[x+i+z, y+j+t]*K[z, t]$
				\EndFor
				\State $score \leftarrow activation\_function(sum)$
				\If {$score > max\_pooling\_element$}
				    \State $max\_pooling\_element \leftarrow score$
				\EndIf
			\EndFor
			\State write $max\_pooling\_element$ to output line buffer
		\EndFor
	\end{algorithmic} 
\end{algorithm}

\begin{figure}[t]
\begin{center}
   \includegraphics[scale=0.6]{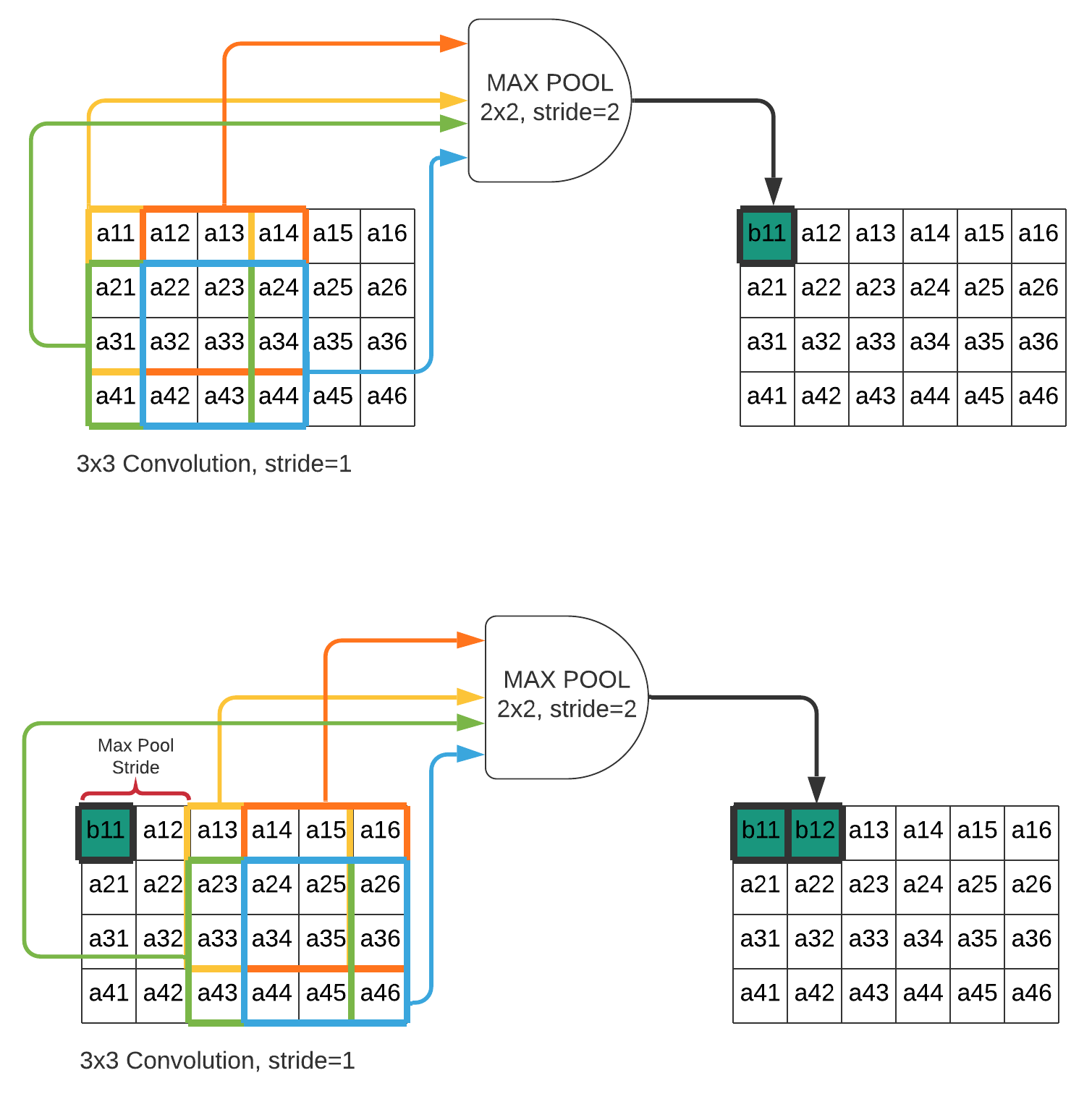}
\end{center}
   \caption{Fused max-pooling layer using reduced dimensions as maximum memory at the output}
\label{fig:long}
\label{fig:onecol}
\end{figure}

This optimization also can reuse input buffer. This may help in different designs but definitely not saving any memory, because maximum memory usage is depending on input size and output size. The greater one determines the maximum memory usage. If we use just this optimization, total memory will be reduced significantly. It was $32*32(input)+6*28*28(conv1)+6*14*14(maxpool1)+16*10*10(conv2)+16*5*5(maxpool2)+120(fc1)+84(fc2)+10(output)=9118*sizeof(float)=\textbf{36472}$ bytes. We can remove all $conv$ layers' output buffers. So it is reduced to $32*32(input)+\stkout{6*28*28(conv1)}+6*14*14(maxpool1)+\stkout{16*10*10(conv2)}+16*5*5(maxpool2)+120(fc1)+84(fc2)+10(output)=2814*sizeof(float)=\textbf{11256}$ bytes. There is \textbf{\%69} memory savings in this example architecture.

\subsection{Using ping-pong buffers between layer outputs}
Since the layer operations are sequential, we do not use whole allocated memory at the same time. This assumption is only valid for single core architectures which covers majority of microcontrollers. In these architectures, parallel operations cannot be performed to make execution pipelined. In the execution, only active layer's input and output are used in each layer operation. Given this fact, total memory usage can be reduced two large line buffers that are exchanging each layer. When next layer is being executed, the output buffer for previous layer is going to be input and previous layer's input will be output buffer. The simple layer illustration is shown in Figure 3. We can reduce total memory usage to sum of maximum two layers' output size. When ping-pong buffers are set max-2 elements in the output buffers, maximum output buffer should be placed first in the layer chain, then other ping-pong buffer will be placed. This guarantees that $2^{nd}$ maximum buffer is never placed maximum sized output. The total memory usage is
$max_{1st}(L) + max_{2nd}(L)$ where $L$ is list of layers output buffer size.\\
If we apply the formula for the example network, we get the following memory usage.
Maximum memory is $max_{1st}(32*32(input), \stkout{6*28*28(conv1)}, 6*14*14(maxpool1), \stkout{16*10*10(conv2)}, 16*5*5(maxpool2), 120(fc1), 84(fc2), 10(output)) + max_{2nd}(32*32(input), \stkout{6*28*28(conv1)}, 6*14*14(maxpool1), \stkout{16*10*10(conv2)}, 16*5*5(maxpool2), 120(fc1), 84(fc2), 10(output))=(1024+1176)*sizeof(float)=\textbf{8800}$ bytes. The relative memory savings from fused in place max-pooling is \textbf{\%22} and the total saving with these two optimizations is \textbf{\%76}.
\begin{figure}[t]
\begin{center}
   \includegraphics[scale=0.7]{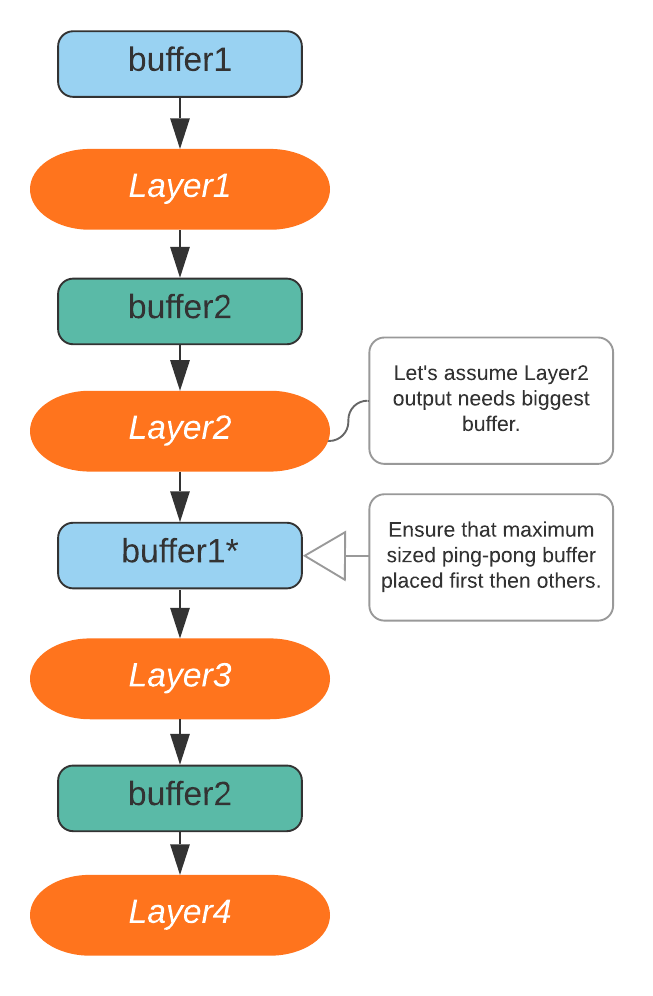}
\end{center}
   \caption{Assignment of ping-pong buffers between layers}
\label{fig:long}
\label{fig:onecol}
\end{figure}

\subsection{Defining parameters read-only}
Parameters are constant and not changed in run time. So they can be defined as read-only. Compiler puts read-only variables into \textit{.text} region. So it never allocates SRAM. \textit{.text} regions are stored in flash memory which is relatively slow to SRAM but cache closes that slow read performance. With this optimization, the total memory usage is only ping-pong buffers and local variables.

\section{Results on LeNet-5 architecture}
C header file which includes all weights in structural form is generated by python code just after the model is trained. The optimized convolution operation(integrated max pooling and activation function) and dot product for fully connected layer are implemented in C. The weight header generator and operator implementations are here \url{https://github.com/hasanunlu/neural_network_deployment_for_uC}. The generated code is compiled for SiFive FE310-G000 RISC-V ~\cite{sifive} microcontroller and the following results are obtained for the compiled source. The SiFive FE310-G000 resources are 16KByte SRAM and 16MByte flash memory. The microcontroller board and the attached camera are shown in Figure 6.
\begin{verbatim}
ELF name: only_network.elf
(usage in bytes)
.text   .data   .bss    .dec    .hex
283318  2764    12032   298114  48c82
\end{verbatim}
Total memory usage is \textit{.data+.bss} = 14796 bytes which is \textbf{$\sim$14 KBytes}. Total flash usage is \textit{.text+.data} = 283318 bytes which is \textbf{$\sim$283 KBytes}. The results are very close to theoretical calculations. The parameter size is 246824 bytes and we get 282632 bytes which also includes instruction, predefined variables and standard C libraries which is around 26 KBytes. The memory usage expectation was around 8.8 KBytes. It resulted as 14.8 KBytes. However, empty project compilation reserves 6 KBytes for stack and heap allocation. If we add them up, the allocation matches with the theoretical  calculations. The execution performance is measured as \textbf{0.26 FPS}(frame per second) for 32x32 gray scale(8 bit per pixel) input at 352 MHz core clock. If we look at performance bottleneck in the systems, the primary one is instruction cache miss while fetching new group of weights. Just after previous layer, new layer parameters are being fetched from flash. It causes instruction cache miss because they are new and not available in the cache at that moment. This small latency happens at the beginning of the each layer execution.

\section{Comparison with related work}
Figure 4. shows the test network implemented with related work and the optimizations proposed here. The pytorch layout of the network is in below. Network is quantized to \textit{int8} for both implementation. The total parameter count is $32*3*5*5+16*32*5*5+32*16*5*5+10*512=$\textbf{33120 = 33 KBytes}.

\begin{verbnobox}[\fontsize{7pt}{7pt}\selectfont]
(0): Conv2d(3, 32, kernel_size=(5, 5), stride=(1, 1),
padding=(2, 2))
(1): ReLU()
(2): MaxPool2d(kernel_size=2, stride=2)
(3): Conv2d(32, 16, kernel_size=(5, 5), stride=(1, 1),
padding=(2, 2))
(4): ReLU()
(5): MaxPool2d(kernel_size=2, stride=2)
(6): Conv2d(16, 32, kernel_size=(5, 5), stride=(1, 1),
padding=(2, 2))
(7): ReLU()
(8): MaxPool2d(kernel_size=2, stride=2)
(9): Flatten()
(10): Linear(in_features=512, out_features=10, bias=True)
\end{verbnobox}

The compilation result using CMSIS-NN framework to deploy this test network is:
\begin{verbatim}
ELF file size: 
Code=10080
RO-data=272
RW-data=36304
ZI-data=48616

Total RAM Size = RW Data + ZI Data
= 85 KBytes

Total ROM Size = Code + RO Data + RW Data
= 46 KBytes
\end{verbatim}

ARM compiler ELF(Executable and Linkable Format) file result is little bit different than RISC-V compiler. All weights are loaded into memory in here. This is significant advantage for execution performance. Using SRAM for constant should be carefully considered. Despite the fact that weights are placed in SRAM in their \cite{lai2018cmsisnn} example, we will still compare memory usage after the weights. In here, it corresponds \textbf{ZI Data = 48 KBytes}. ARM architecture also uses reserved stack and heap area which is 4 KBytes in this example. The corrected utilization for RAM would be 48 KBytes - 4 KBytes = 44 KBytes. ROM utilization for weights is directly equal to \textbf{RW Data = 36 KBytes}.

\begin{figure}[t]
\begin{center}
   \includegraphics[scale=0.47]{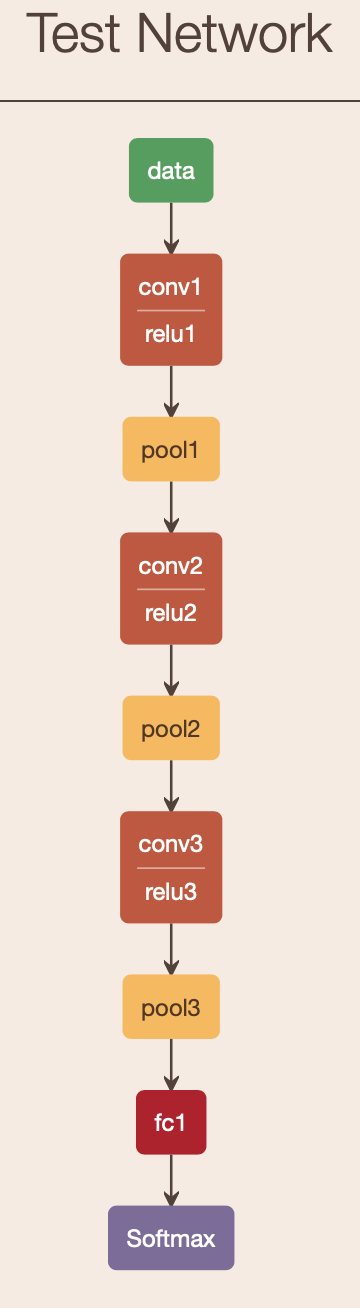}
\end{center}
   \caption{Test network for memory comparison}
\label{fig:long}
\label{fig:onecol}
\end{figure}

The result with these two optimization technique for RISC-V architecture is shown in below.
\vspace*{-12mm}
\begin{verbatim}
ELF name: test_network.elf
(usage in bytes)
.text  .data   .bss    .dec    .hex
66678  2764    14500   83942   147e6
\end{verbatim}
\vspace*{-10mm}

In section 4, total stack and heap allocation indicated 6 KBytes. The corrected RAM utilization would be \textbf{11.2 KBytes}. ROM utilization for weights in RISC-V compiler will be calculated from \textit{.text} size. But it has code region as well. The empty project consumes 26 KBytes. The code size is measured as 4 KBytes. The remaining \textit{.text} size is equal to weights initialization after these two allocations. Weights utilization in ROM is 66 KBytes - 4 KBytes - 26 KBytes = \textbf{36 KBytes}.

\begin{table}[hbt!] 
\begin{tabular}{||c c c c||} 
\hline
Region & CMSIS-NN \cite{lai2018cmsisnn} & Our framework & Difference\\ [0.5ex] 
\hline\hline
ROM & 36 KBytes & 36 KBytes & \%0 \\ 
\hline
RAM & 44 KBytes & \textbf{11.2 KBytes} & \%74 less \\
\hline
\end{tabular}
\vspace*{2mm}
\caption{Corrected RAM and ROM utilization for the test network.}
\end{table}

As we see, ROM utilization is same for both compiled sources. The size is also very close to theoretical parameter count calculation (33 KBytes). However, RAM utilization is significantly less in our framework. The main advantage is coming from fused max-pooling layer at the end of convolutions. It reduces the total memory consumption to $m*n/s^2$ instead of $m*n$. Fine tuned calculation of ping-pong buffer approach also provides some savings. Because, CMSIS-NN \cite{lai2018cmsisnn} uses maximum of the output size of the layers as scratch line buffers. Our approach limits maximum ping-pong buffer size using maximum first two elements of the output sizes.

\section{Image pipeline and LeNet-5 Demo}
In section 4, LeNet-5 is trained on MNIST and deployed to RISC-V microcontroller board. The platform shown in Figure 6 is used to be able to show real time demo using cameras and the RISC-V microcontroller. The camera is capable of capturing 2 Megapixel RGB(red/green/blue) images and it has configurable hardware FIFO(first in first out) for cropped resolutions.

\begin{figure}[H]
\begin{center}
   \includegraphics[scale=0.6]{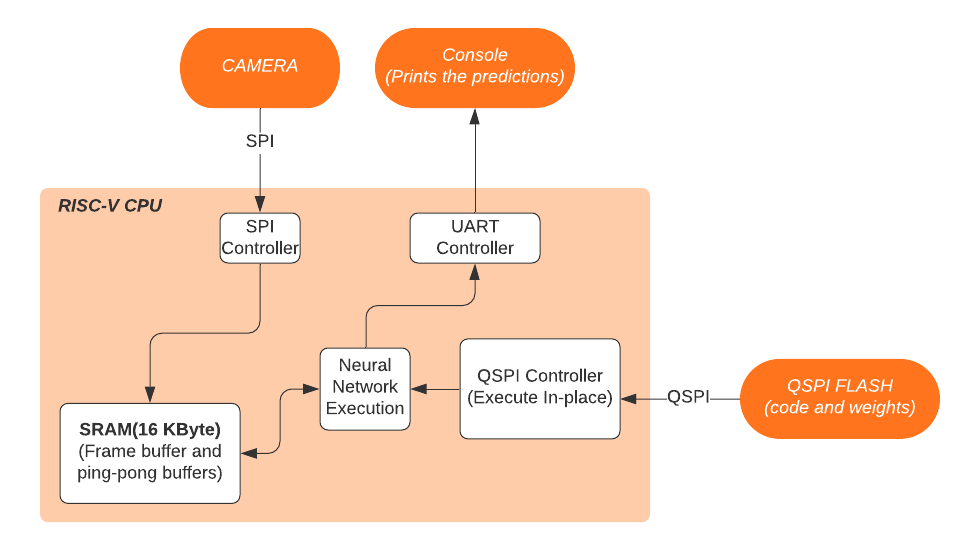}
\end{center}
   \caption{Data flow and hardware connections}
\label{fig:long}
\label{fig:onecol}
\end{figure}

The resolution is scaled to 32x32 and gray scale(8 bits per pixel) color option is selected. Every manual trigger from microcontroller dumps image frame to hardware FIFO. Then, microcontroller reads FIFO using SPI(Serial Peripheral Interface). The image is stored into frame buffer to be processed later. MNIST data set is white on black background. The background is pure black, so we have to make camera captured images very similar to this form. While storing the image, it is inverted($255-pixel\_value$) and, if the pixel value is smaller than pixel value of 100, it is recorded as pure black(pixel value is 0). This basic filtering ensembles camera input data to MNIST dataset texture. Figure 5 shows data pipeline for processing flow. Frame buffer pointer is delivered to neural network's input and processing starts. After execution is done, the best scored class is printed on UART(Universal Asynchronous Receiver/Transmitter) console.

\begin{figure}[H]
\begin{center}
   \includegraphics[scale=0.101]{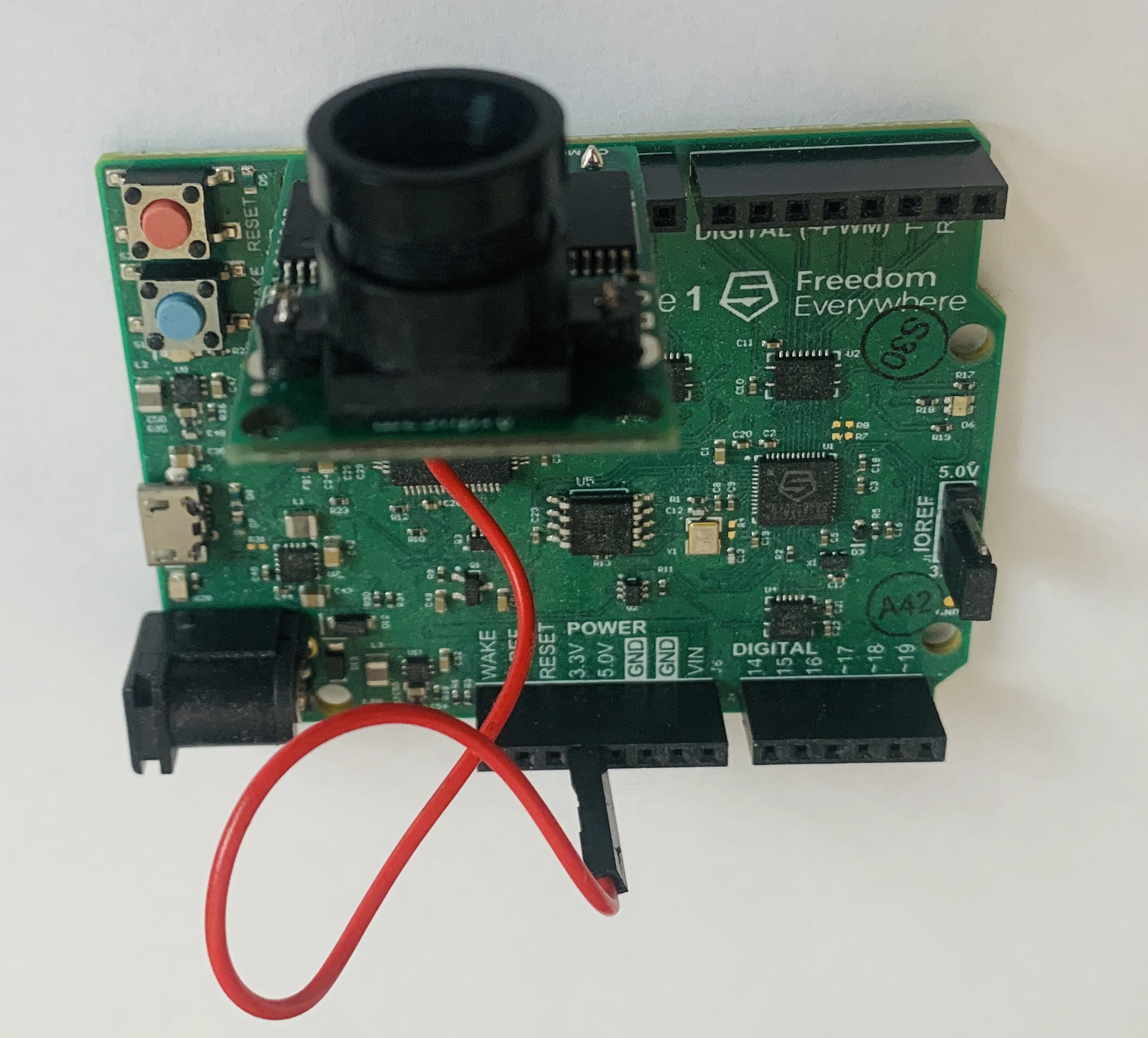}
\end{center}
   \caption{SiFive FE310-G000 RISC-V with OV2640 camera}
\label{fig:long}
\label{fig:onecol}
\end{figure}

\section{Future Work}
This paper shows neural networks can be deployed microcontrollers using significantly less memory than the original size. The primary optimization allowing us to make this significant memory reduction is the fused in-place max-pooling. It currently requires stride is greater than or equal to pooling kernel size to make output buffer optimization possible. $stride>=k=k$ actually covers all pooling methods used in practice now. As a next step, I will augment max-pooling optimization to any stride using some memory (which is less than or equal to pooling kernel size) while convolution is being calculated. This paper is mainly focused on memory optimizations. Convolutions and any matrix products can be accelerated by SIMD(single instruction multiple data) instruction in different platforms. In the tool, depending on remaining RAM resource, some weights can be moved into RAM, so it makes execution faster for those weights, because there is no longer any need to fetch weights from flash(slower than RAM) memory. This approach is convenient for convolution kernel weights. They are small and repetitively used in same data input.

{\small
\bibliographystyle{unsrt}
\bibliography{egbib}
}

\end{document}